\documentclass[usegraphicx]{XrU2005}
\usepackage{graphicx}
\usepackage{rotating}
\usepackage{amsmath}

\def\lesssim{\mathrel{\hbox{\rlap{\hbox{\lower4pt\hbox{$\sim$}}}\hbox{$<$}}}}

\title{An explanation for the soft X-ray excess in AGN}
\author{J. Crummy}
\author{A.C. Fabian}
\affil{Institute of Astronomy, Madingley Road, Cambridge, UK. CB3 0HA}
\author{L. Gallo}
\affil{Max-Planck-Institut f\"ur extraterrestrische Physik, Postfach 1312, 85741 Garching, Germany}
\author{R.R. Ross}
\affil{Physics Department, College of the Holy Cross, Worcester, MA 01610, USA}

\begin{document}

% keywords
\keywords{accretion; accretion discs; active galactic nuclei; X-rays}

\maketitle

\begin{abstract}
We present a large sample of type 1 AGN spectra taken with XMM-Newton. We fit them with the relativistically blurred photoionized disc model of Ross \& Fabian (2005). This model is based on an illuminated accretion disc of fluorescing and Compton-scattering gas, and includes relativistic Doppler effects due to the rapid motion of the disc and general relativistic effects such as gravitational redshift due to presence of the black hole. The disc model successfully reproduces the X-ray continuum shape, including the soft excess, of all the sources. It provides a natural explanation for the observation that the soft excess is at a constant temperature over a wide range of AGN properties. The model also reproduces many features that would conventionally be interpreted as absorption edges. We use the model to measure properties of the AGN such as inclination, black hole rotation, and metallicity.
\end{abstract}
\section{Introduction}
The soft excess is an important component of the X-ray spectra of many AGN and is present in every source in this survey. The soft excess is defined as the enhanced emission below $\sim$2~keV compared to an extrapolation of the approximately power law spectrum in the 2 -- 10~keV band. This extra emission is generally approximately thermal in shape, and well fit with a black body of energy 0.1 -- 0.2~keV. This result stands over several decades in AGN mass, e.g. Walter \& Fink (1993),  Gierli\'{n}ski \& Done (2004), Porquet et al. (2004). This poses problems for the thermal interpretation of the data, as the temperature does not scale in the expected way with luminosity ($L \propto T^4$), as well as being too high for the standard model of Shakura \& Sunyaev (1973). Several alternative models have been proposed to account for one or both of these results, none of which have yet been broadly accepted by the community. This paper investigates photoionized emission blurred relativistically by motion in an accretion disc, which has been previously studied by e.g. Ballantyne, Iwasawa \& Fabian (2001). We use the latest models from Ross \& Fabian (2005), which include more ionization states and more recent atomic data than earlier versions (Ross \& Fabian 1993).\\
\section{The relativistically blurred photoionized disc model}
\begin{figure*}[t]
\centering
\includegraphics[angle=270, width=0.425\linewidth]{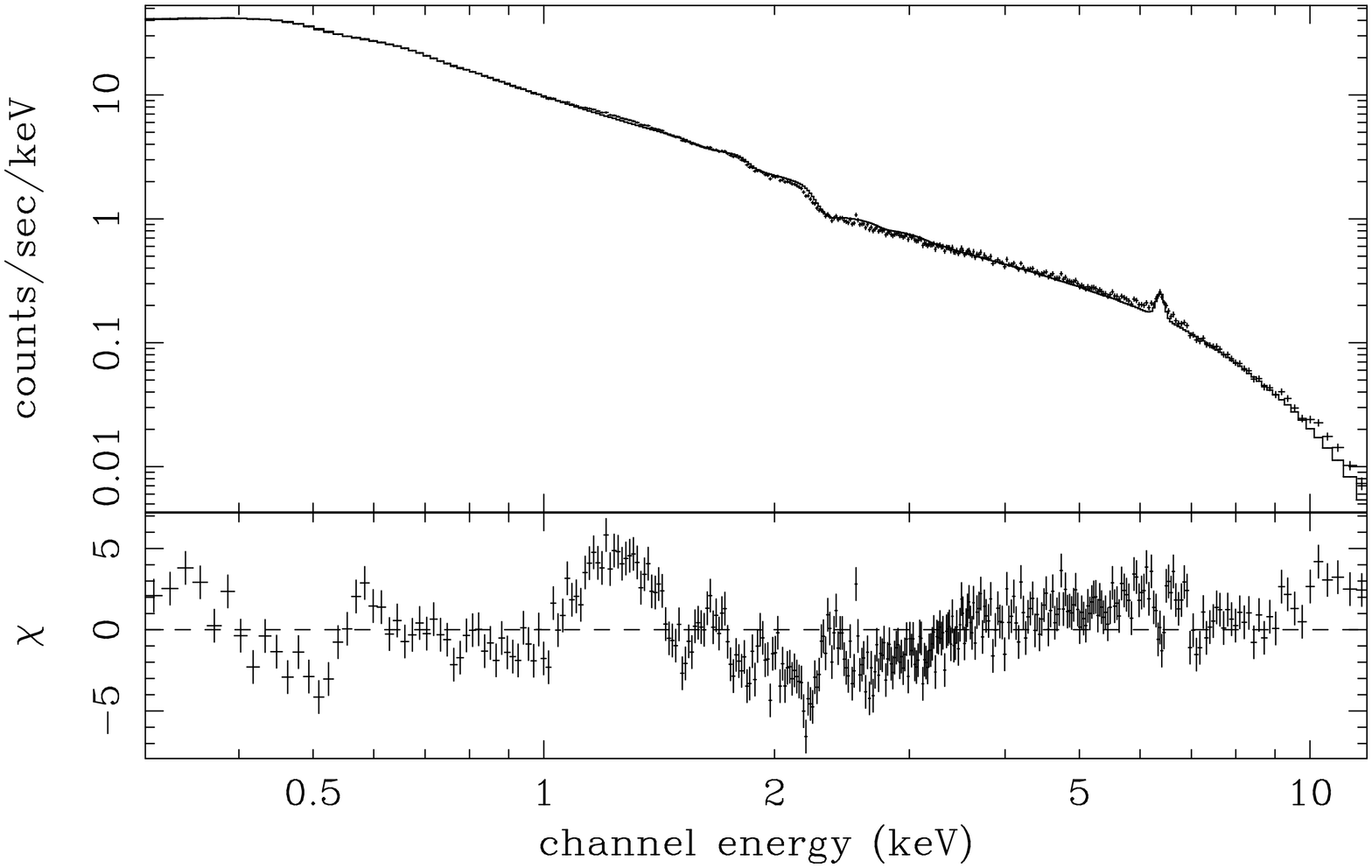}
\hspace{11mm}
\includegraphics[angle=270, width=0.425\linewidth]{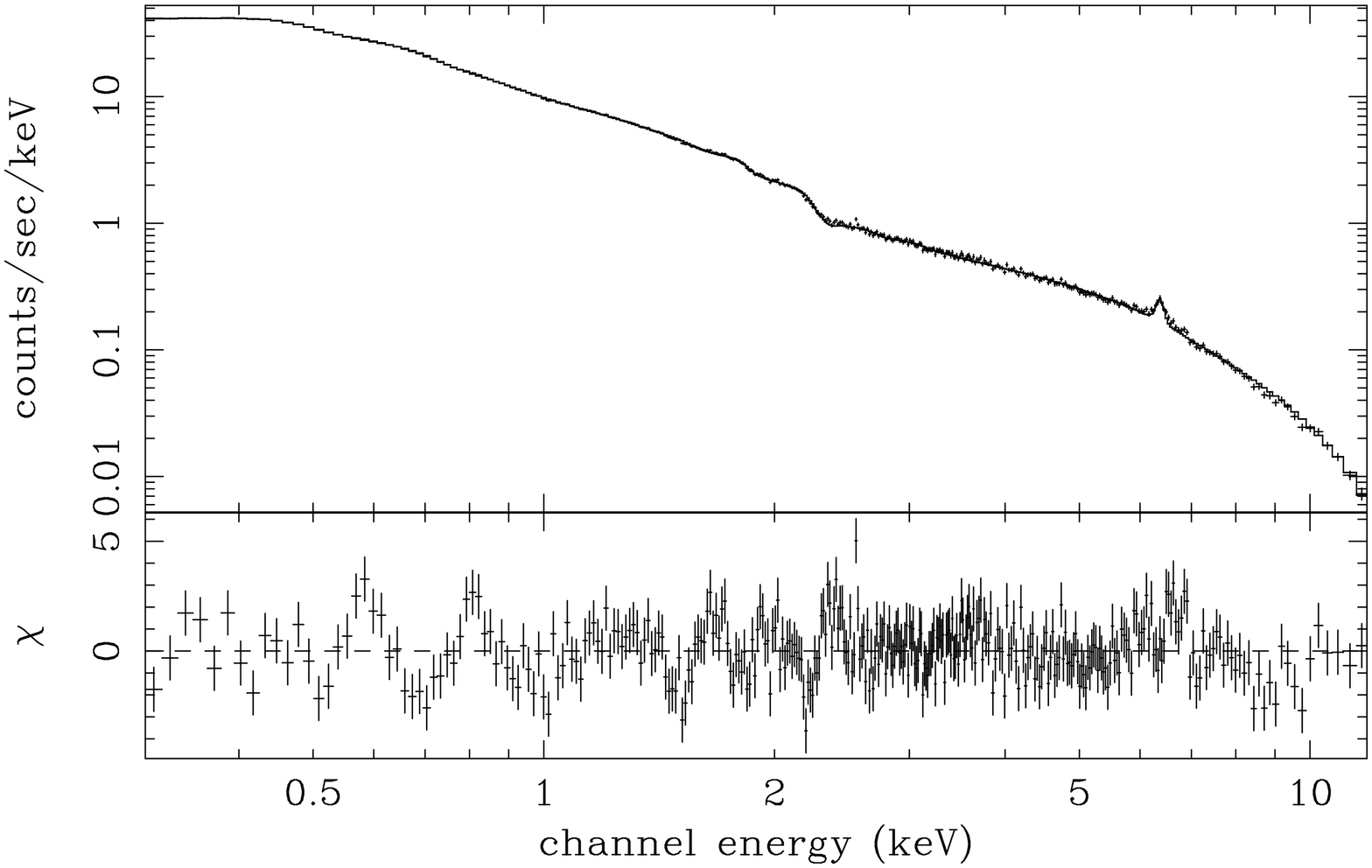}
\includegraphics[angle=270, width=0.425\linewidth]{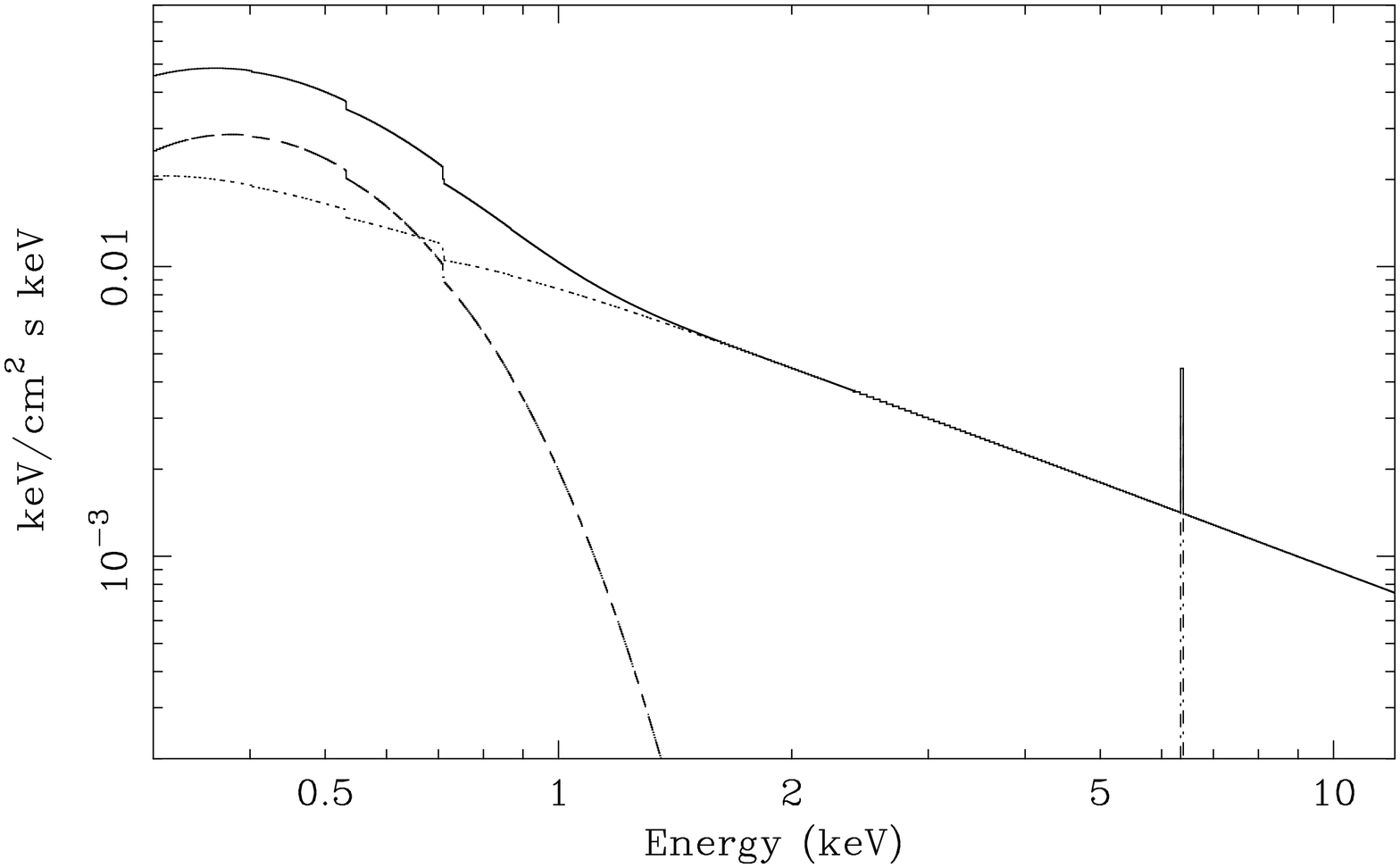}
\hspace{11mm}
\includegraphics[angle=270, width=0.425\linewidth]{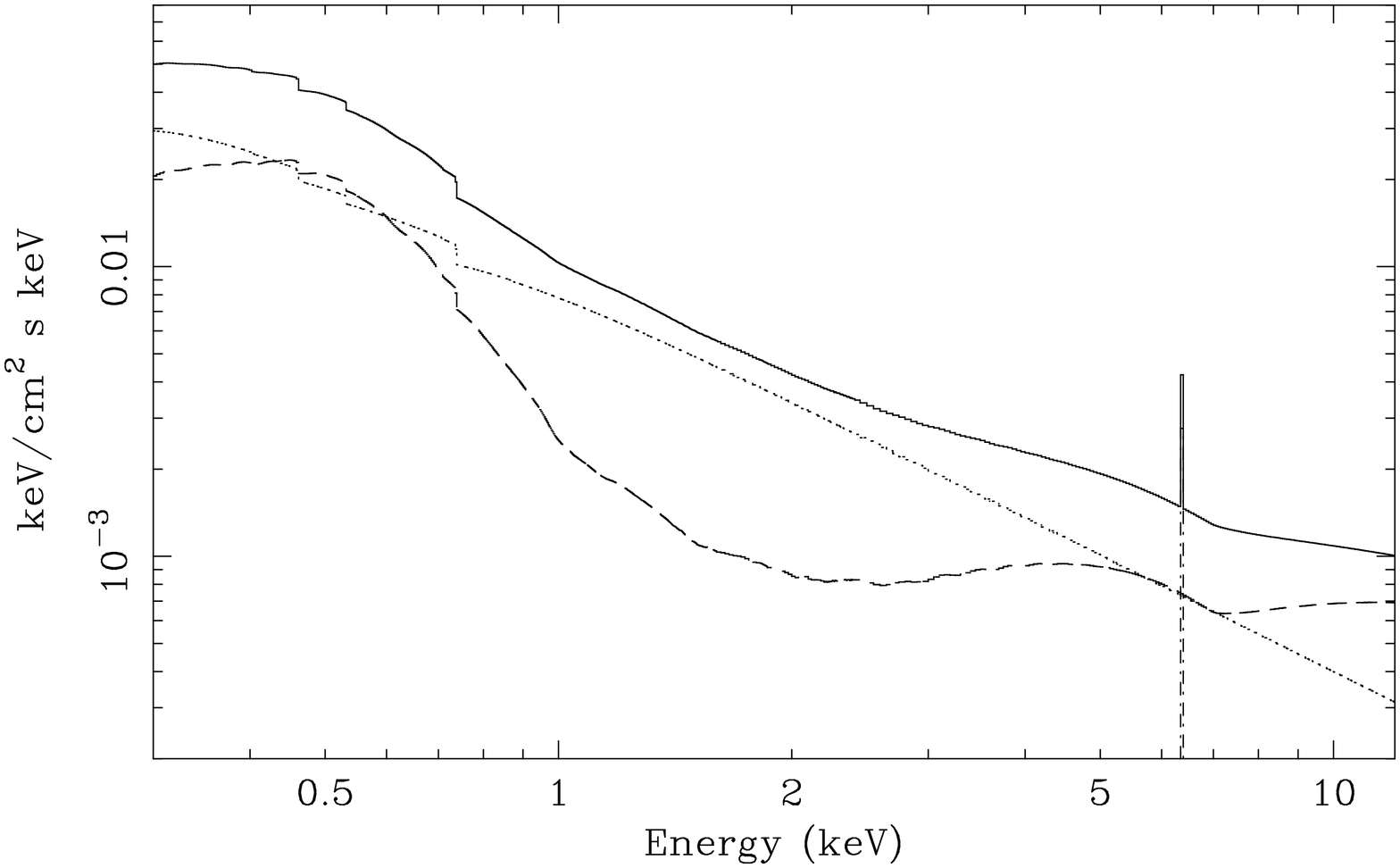}
\caption{This figure shows a comparison between the thermal model and the relativistically blurred photoionized disc reflection model for NGC 4051. The top panels show spectra and residuals to the thermal (left) and disc reflection (right) models, and the bottom panels show the components of the thermal (left) and disc reflection (right) models. Both include a power law (dotted line) and narrow iron emission (sharp line at $\sim$6.4~keV), the components shown in dashed lines are a black body and the relativistically blurred disc reflection respectively. As the residuals show, the disc reflection model is a much better fit. Most of the remaining residuals to NGC 4051 in the soft band (top right panel, 0.5 -- 1.0~keV) have been shown to be due to narrow line emission (Ponti et al., in prep).\label{4051_figure}}
\end{figure*}

In the model of Ross \& Fabian (2005) a semi-infinite slab of cold optically thick gas of constant density is illuminated by a power law, producing a Compton component and fluorescence lines from the ionized species in the gas. To produce the relativistically blurred photoionized disc model this reflected emission is added to the illuminating power law and the summed emission is convolved with a Laor (1991) profile to simulate the blurring from an accretion disc around a maximally rotating (Kerr) black hole. This assumes a geometrically thin and flat accretion disc with clearly defined inner and outer radii, and that the emissivity of the disc as a function of radius is described by a power law. This combined model allows fitting to the inclination of the disc, the inner and outer radii (the outer radius is generally poorly constrained as emission is centrally concentrated), the emissivity index of the disc, the iron abundance of the gas, the spectral index of the illuminating power law and the ionization parameter (which we define as $\xi = 4 \pi F/n_{H}$, where $F$ is the illuminating energy flux and $n_{H}$ is the hydrogen number density in the illuminated layer, a measure of the ratio of the energy density in the illuminating radiation to the atomic number density in the gas). Since the model is physically motivated measuring these free parameters can give us information about the sources. An example of the model is shown in Figure \ref{4051_figure} (bottom right panel).\\
\section{Data}
We used publicly available archival \textit{XMM-Newton} data on 22 type 1 AGN from the Palomar-Green (PG) sample and a selection of 12 other Seyfert 1 galaxies with high-quality observations available. The sample includes several well-studied AGN, such as NGC 4051, PG 1211+143 and I Zw 001. We analyse the longest available observation where the EPIC \texttt{pn} camera took data. We reduced the Observation Data Files in the standard way using \texttt{SAS 6.0} to produce spectra (See Crummy et al. 2005b for details), taking the range 0.3 -- 12.0~keV as the region in which the \texttt{pn} is accurately calibrated. We do not include \texttt{MOS} data in our fits. We grouped the spectra so each bin includes at least 20 source counts, so $\chi^{2}$ statistics are applicable. Quoted errors are 90 per cent limits on one parameter ($\Delta \chi^{2}$ = 2.706).\\
\section{Analysis}
We fit the data with two classes of model using \texttt{xspec}, the standard power law with a black body to model the soft excess, and the relativistically blurred photoionized disc reflection model. Both models are subject to absorption from cold gas in our Galaxy, we fix the amount of absorption at the value given by the \texttt{nh} ftool (Dickey \& Lockman 1990). We also allow for absorbing matter at the AGN by fitting for extra cold absorption and up to two absorption edges in the 0.45 -- 1.1~keV band. These components are redshifted such that they are local to the AGN. We finally allow for cold reflection (e.g. from distant gas such as a torus) by including a narrow iron line at 6.4~keV (in the source frame). We are concerned primarily with the underlying continuum rather than these features, so we include in our final model only those components which improve the overall $\chi^{2}$ by $>$ 2.7 per lost degree of freedom. Since we are performing fits over the entire \textit{XMM-Newton} band, we only report lines for which \texttt{xspec} reports a non-zero minimum equivalent width, this avoids fitting spurious lines due to a curvature in the continuum not addressed by the model. This conservative procedure means faint lines are likely to be missed, although any line that significantly affects the overall goodness of fit will be included. In xspec terminology our models are \texttt{phabs * zphabs * zedge * zedge * (powerlaw + zbbody + zgauss} and \texttt{phabs * zphabs * zedge * zedge * (kdblur(powerlaw + atable\{reflion\}) + zgauss)}, where  \texttt{kdblur} is a convolution model using a Laor line and  \texttt{reflion} is a table model of reflection from cold gas (including a redshift)\footnote{Models available from\\ http://www-xray.ast.cam.ac.uk/$\sim$jc/kdblur.html and\\ http://heasarc.gsfc.nasa.gov/docs/xanadu/xspec/models/reflion.html}.\\
\begin{figure*}[t]
\centering
\includegraphics[angle=270, width=0.425\linewidth]{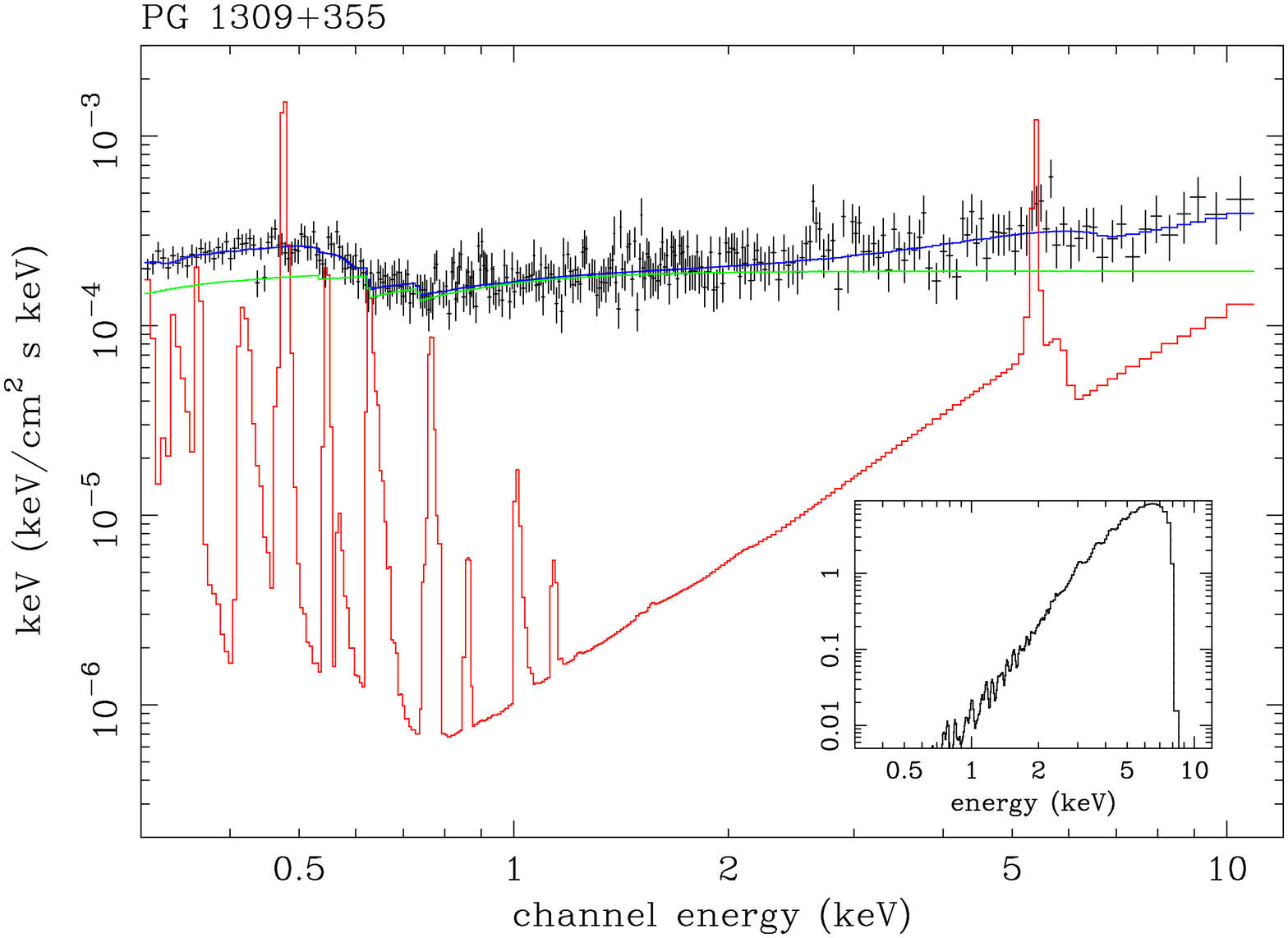}
\hspace{11mm}
\includegraphics[angle=270, width=0.425\linewidth]{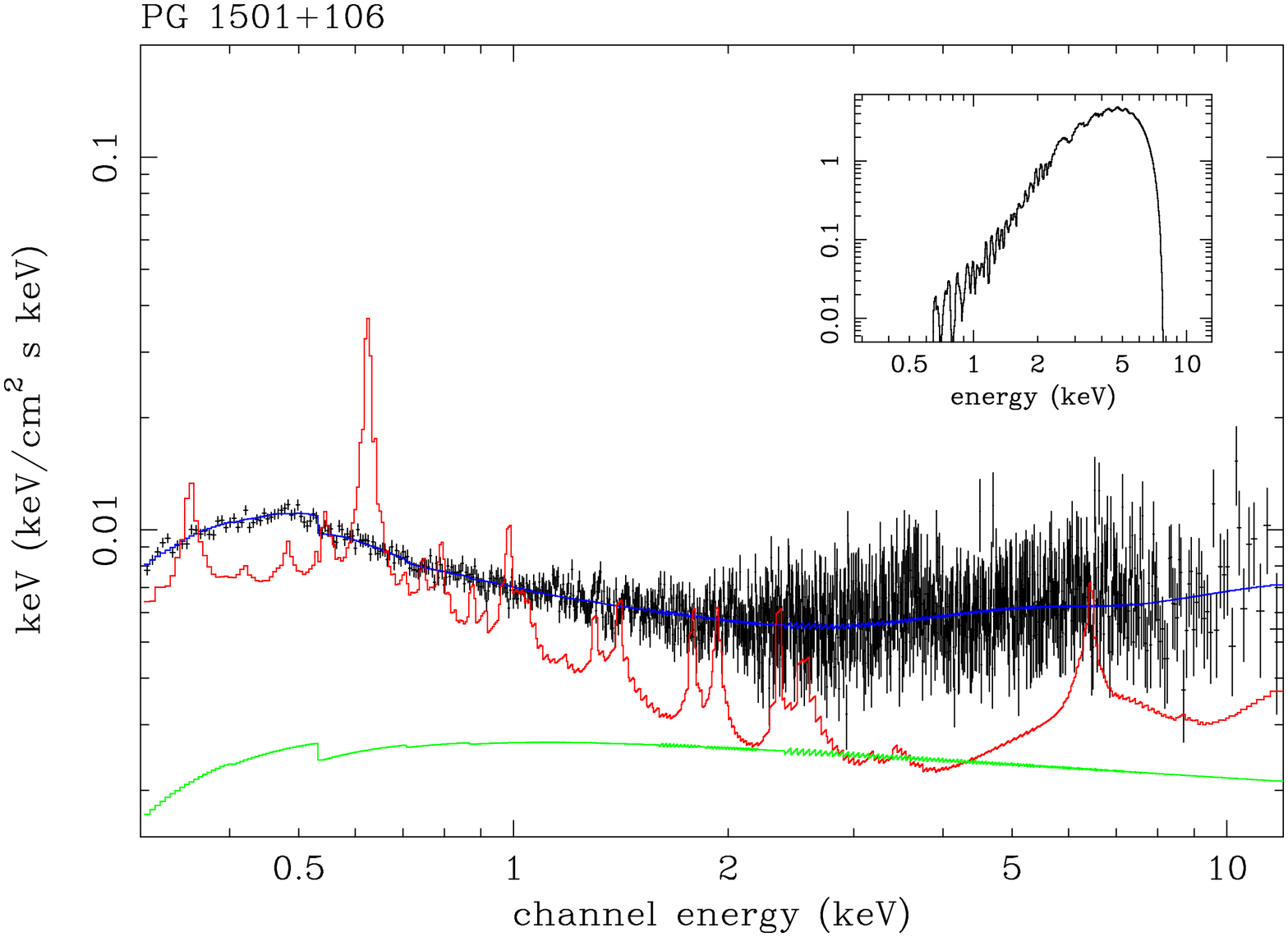}
\includegraphics[angle=270, width=0.425\linewidth]{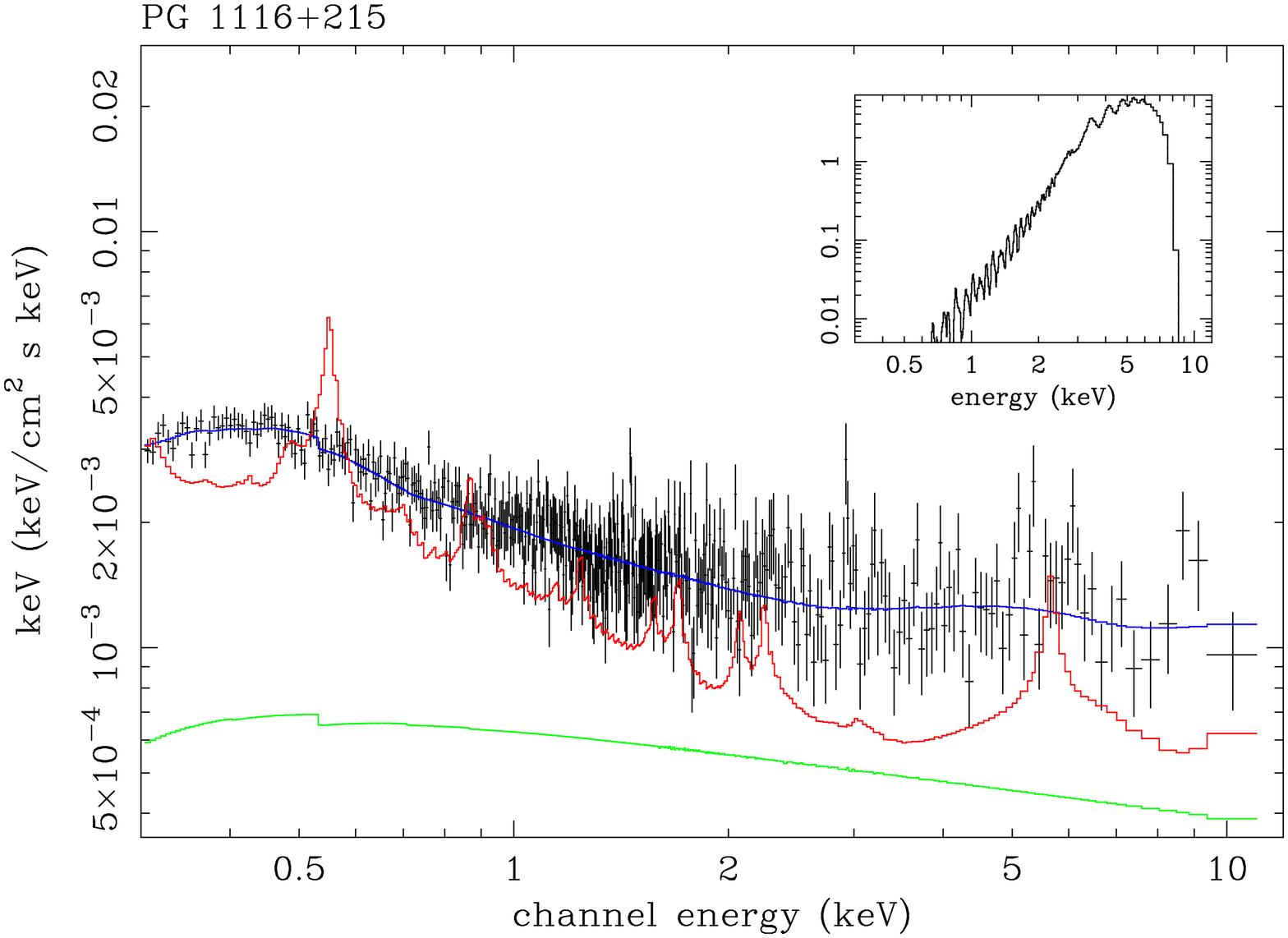}
\hspace{11mm}
\includegraphics[angle=270, width=0.425\linewidth]{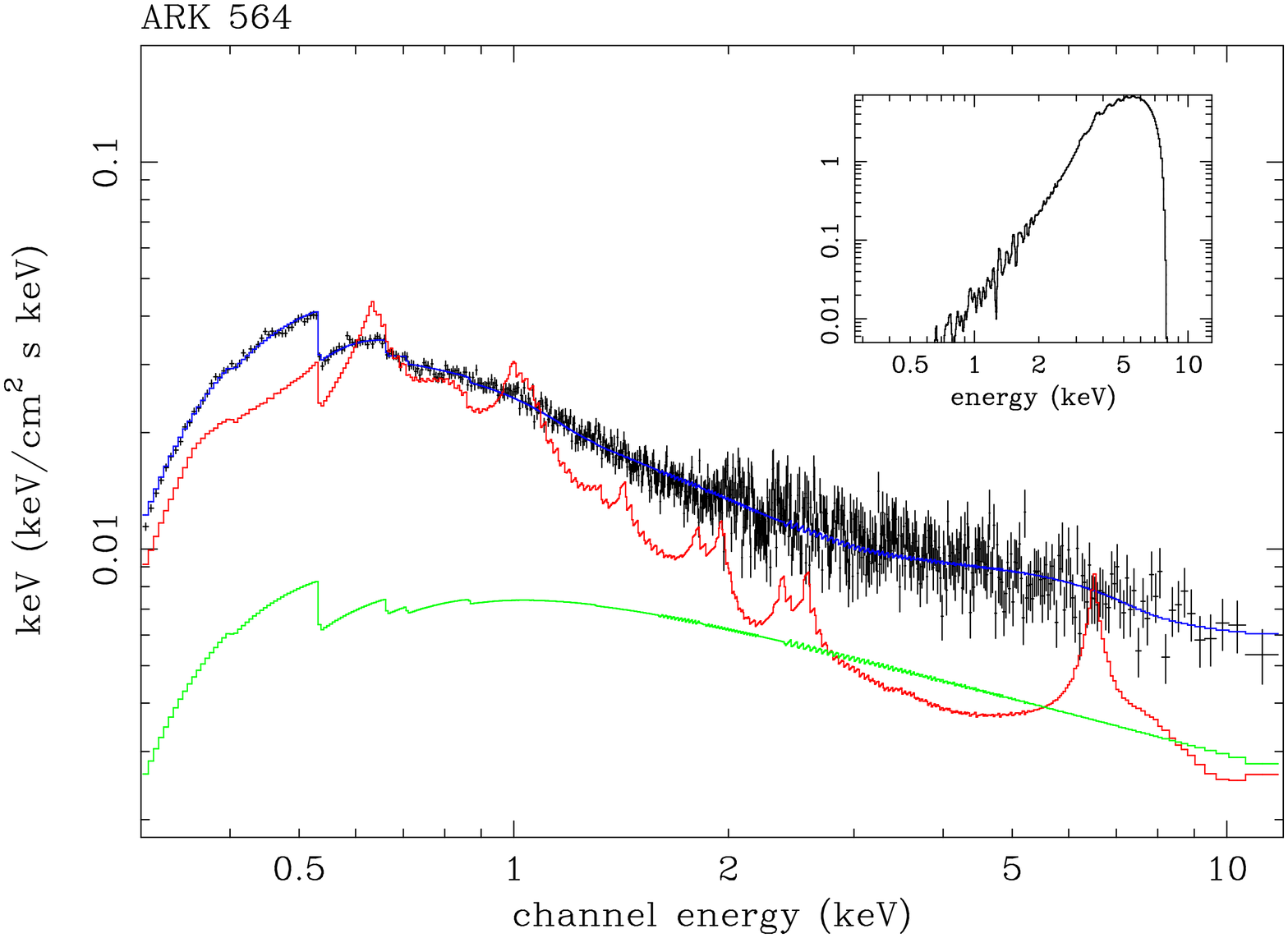}
\caption{Example disc reflection model fits plotted in $\nu F_{\nu}$, showing the amount of flux emitted as a function of energy. The crosses are data, the data-following (blue) line is the fit, the smooth (green) line the power law component and the spiky (red) line the reflection component. The inserts show the shape of the Laor line used in the convolution. The soft excess is shown to be the result of blurred line emission, and the extreme blurring effect on the iron lines around 6.4~keV is visible. The figure also illustrates the effect of ionization parameter on reflection spectra, for PG 1309+355, $\xi =$ 3, PG 1501+108, $\xi =$ 510, PG 1116+215, $\xi =$ 1270 and ARK 564, $\xi =$ 3120. A colour version is available online.\label{nufnu_figure}}
\end{figure*}
\section{Results \& Discussion}
Our results are presented in full in Crummy et al. 2005b. One of the sources in the sample, PG 1404+226, has been investigated with the same model in a previous paper (Crummy et al. 2005a). We find that in general the relativistically blurred photoionized disc reflection model is a better fit to the data, with 25 of the 34 sources showing an improvement in $\chi^{2} >$ 2.7 per degree of freedom (note that this does not correspond to 90 per cent probability, the $\chi^{2}$ distribution is not calibrated across models). 6 sources show a significant worsening and 3 sources are inconclusive. In some sources the improvement is very marked, see Figure \ref{4051_figure}.\\
The disc reflection model reproduces the shape of the continuum well, and naturally explains the constant temperature of the soft excess; see Figure \ref{nufnu_figure}. The model also reproduces many features that would be otherwise interpreted as absorption edges in medium resolution \texttt{pn} data. When the thermal model is used, 17 of the 34 sources appear to have an absorption edge. This reduces to just 7 sources when the disc reflection model is used. The disc reflection model is somewhat less smooth than the thermal model (see Figure \ref{4051_figure}), and bumps on the scale of absorption edges are possible. Absorption lines cannot be reproduced by the model, therefore the possibility of outflows from absorption edges should be checked by investigating reflection models, or by detecting absorption lines from outflowing matter. This has implications for the energetics and cosmological evolution of these systems, as some measured outflows can involve similar amounts of power as the radiation e.g. Pounds et al. (2003). The model also explains the apparent lack of broad iron lines in most AGN spectra. Very few broad iron lines have been clearly detected (the most well studied being MCG --6-30-15, Tanaka et al. 1995), which is unexpected since it is believed that these systems are powered by accreting matter near the black hole, where broad iron lines are formed. Our results suggest that the broad iron lines do occur, but are broadened to the point of near undetectability. A previous paper by Gallo et al. (2004) supports this hypothesis, they fit an extremely broad excess of counts in the hard band of MRK 0586 with a Laor profile. MRK 0586 is included in our sample, and we find an inclination consistent with theirs; we even fit consistent inclinations when we exclude the hard band from the fit and determine the inclination purely from the blurring effects on the soft excess. This illustrates how the hard and soft bands are connected, they key to the absence of obvious broad lines is in the shape of the soft excess. Conversely, it is clear that the soft excess is nothing more than a blend of many broad lines.\\
\begin{figure}
\centering
\includegraphics[width=0.85\linewidth]{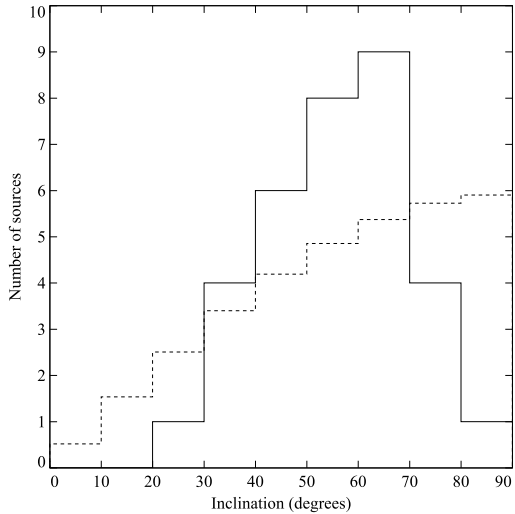}
\caption{This histogram shows the measured inclinations of the sources (the solid line) and the expected distribution if the inclinations were random (the dashed line). There is a deficit at high inclinations.\label{inclinations_figure}}
\end{figure}
As noted in the previous paragraph, it is also possible to measure the inclination of the central disc using the disc reflection model. Our measured inclinations are plotted in Figure \ref{inclinations_figure}. We find that the inclinations are inconsistent with being random, with a Kolmogorov-Smirnov (K-S) test probability of 0.06. Performing a K-S test against a distribution which is random over $0^{\circ}$ -- $81^{\circ}$ (ignoring any data points above  $81^{\circ}$) gives a probability of 0.34, so the data is somewhat consistent with being random over this range. The deficit at low inclinations may be a selection effect, discs at $\sim$$0^{\circ}$ Doppler beam all of their emission along the plane of the disc and so would appear dim. The model also takes no account of limb darkening, which is strong at low inclinations. At high inclinations the deficit could be due to torus obscuration, with the sources that do have high inclination perhaps not having a torus, or with the torus unaligned with the central disc.\\
Abundances may also be measured with the model. The model assumes a fixed, solar abundance for all elements except iron, for which the abundance is a free parameter. The measured abundance is consistent with being between solar and $1/3$ solar for 28 of the 34 sources. No sources have measured abundances below $1/3$ solar, and 6 have abundances which are clearly above solar. These fairly low iron abundances are another factor in the apparent absence of broad iron lines. The inability to fit more than one element abundance means these results are tentative, but show that in principle the composition of AGN accretion discs can be measured from their X-ray spectra.\\
Finally, the rotation of the central black hole may be measured using this model. The model uses convolution with a Laor line, which is based on a maximally rotating (Kerr) black hole. It is equally possible to use a model based on a non-rotating (Schwarzschild) black hole (\texttt{diskline} in \texttt{xspec}, Fabian et al. 1989). We also performed fits to all our sources using this non-rotating model, and found that in all cases the non-rotating model has a worse goodness of fit, with only two of the 34 sources where the fit is of comparable quality. This shows that rotating black holes dominate our sample. The black hole rotation is not a free parameter in our analysis, but the inner radius of the accretion disc is. The last stable orbit around a non-rotating black hole is at 6 gravitational radii, and the last stable orbit around a maximally rotating black hole is at 1.235 gravitational radii. Inside this radius matter plunges into the black hole, and does not emit strongly (Section 3.4 of Fabian \& Miniutti, 2005). All 34 sources have inner disc radii consistent with being below 6 gravitational radii, and 29 of the 34 have inner radii consistent with being below 1.3 gravitational radii. This is a strong indication that most black holes are maximally rotating, in agreement with theoretical predictions, e.g. Volonteri et al. (2005) predicts that 70 per cent of AGN black holes are maximally rotating.\\
One further interesting result from the fits is that the illuminating power-law component is often undetected. This seems to be a problem, as it is hard to imagine how the disc could be illuminated by radiation we cannot see. However, it is quite possible for relativistic effects due to the black hole to ``hide'' the illuminating continuum from us. If the source of the illuminating continuum is above the disc, the light from it will be bent and impact on the disc, with little escaping to be observed (Miniutti \& Fabian 2004). Emission from the disc is Doppler beamed away from the black hole and will still escape. This primarily applies to sources which do not share the rotation velocity of the disc, e.g. the base of a weak jet or shocks in a failed jet (Ghisellini, Haardt \& Matt 2004).
\section{Conclusions}
We fit a large number of type 1 AGN with a thermal model and a relativistically blurred photoionized disc reflection model based on Ross \& Fabian (2005). We find that:
\begin{itemize}
\item The disc reflection model fits the data better.
\item The disc reflection model reproduces all the major features of all the sources, including the soft excess and the absence of obvious broad iron lines. The model explains the constant temperature of the soft excess across all the sources, since it is formed from a large number of highly broadened lines. The model reproduces many features that might otherwise be interpreted as absorption edges.
\item Black holes in AGN strongly rotate.
\item The central discs around type 1 AGN have a wide distribution in inclination, with possible deficits at very low and high inclinations. The high inclination deficit may indicate torus obscuration.
\item The elemental abundances in accretion discs can be measured, iron abundances in these sources are solar or mildly sub-solar.
\end{itemize}
The relativistically blurred photoionized disc reflection model is an important tool in the study of AGN. Taking account of their intrinsically relativistic nature answers several questions about their spectra, as well as providing information about the central regions.
\section*{Acknowledgments}
JC, ACF and RRR thank PPARC, the Royal Society, and the College of the Holy Cross for support respectively. The \textit{XMM-Newton} satellite is an ESA science mission (with instruments and contributions from NASA and ESA member states).
\section*{References}
Ballantyne D.R., Iwasawa K., Fabian A.C., 2001, MNRAS, 323, 506\\
% fits from early disc reflection model
Crummy J., Fabian A.C., Brandt W.N., Boller Th., 2005a, MNRAS, 361, 1197\\
%PG 1404
Crummy J., Fabian A.C., Gallo L., Ross R.R., 2005b, MNRAS accepted, astro-ph:0511457\\
%Soft excess paper
Dickey J.M., Lockman F.J., 1990, ARA\&A, 28, 215\\
%galactic hydrogen
Fabian A.C., Miniutti G., 2005, astro-ph:0507409\\
% review chapter
Fabian A.C., Rees M.J., Stella L., White N.E., 1989, MNRAS, 238, 729\\
%diskline
Gallo L.C., Boller Th., Brandt W.N., Fabian A.C., Vaughan S., 2004b, MNRAS, 355, 330\\
% MRK 0586 very broad line
Ghisellini G., Haardt F., Matt G., 2004, A\&A, 413, 535\\
% aborted jets
Gierli\'{n}ski M., Done C., 2004, MNRAS, 349, L7\\
%(is the soft excess real - absorption...)
Laor A., 1991, ApJ, 376, 90\\
%(Laor profile as in kdblur)
Miniutti G., Fabian A.C., 2004, MNRAS, 349, 1435\\
%(light bending model)
Porquet D., Reeves J.N., O'Brien P., Brinkmann W., 2004, A\&A, 422, 85\\
% XMM obs of 21 PG QSOs
Pounds K.A., Reeves J.N., King A.R., Page K.L., O'Brien P.T., 2003, MNRAS, 345, 705\\
% PG 1211 outflows etc.
Ross R.R., Fabian A.C., 1993, MNRAS, 261, 74\\
% early disc reflection model
Ross R.R., Fabian A.C., 2005, MNRAS, 358, 211\\
%(reflion model)
Shakura N.I., Sunyaev R.A., 1973, A\&A, 24, 337\\
% original accretion disc
Tanaka Y., et al., 1995, Nature, 375, 659\\
%ASCA mcg6, line profile 11 co-authors
Volonteri M., Madau P., Quataert E., Rees M.J., 2005, ApJ, 620, 69\\
% black hole spins
Walter R., Fink H.H., 1993, A\&A, 274, 105\\
% const temp soft excess
Wang T., Lu Y., 2001, A\&A, 377, 52\\
%(variability properties of AGN)
\end{document}